\documentclass[twocolumn,showpacs,preprintnumbers,amsmath,amssymb,superscriptaddress,aps,pra,10pt]{revtex4-1}

\usepackage{color}
\usepackage{graphicx}
\usepackage{citesort}
\usepackage{color}
\usepackage{subfigure}
\usepackage{upgreek}

\newcommand{\comment}[1]{}

\newcommand{\Pop}{P}
\newcommand{\Pac}{P_{\rm ac}}
\newcommand{\vop}{v}
\newcommand{\vac}{v_{\rm ac}}

\begin{document}

\title{Crosstalk-free multi-wavelength coherent light storage via Brillouin interaction}
\author{B.~Stiller}
\thanks{These two authors contributed equally.}

\affiliation{
Institute of Photonics and Optical Science (IPOS), The University of Sydney Nano Institute, School of Physics, University of Sydney, Sydney, NSW, Australia
}
\email{birgit.stiller@sydney.edu.au}

\author{M.~Merklein}
\thanks{These two authors contributed equally.}
\affiliation{
Institute of Photonics and Optical Science (IPOS), The University of Sydney Nano Institute, School of Physics, University of Sydney, Sydney, NSW, Australia
}

\author{K. Vu}
\affiliation{
  Laser Physics Centre, RSPE, Australian National University, Canberra, ACT 2601, Australia
}

\author{P. Ma}
\affiliation{
  Laser Physics Centre, RSPE, Australian National University, Canberra, ACT 2601, Australia
}

\author{S.~J. Madden}
\affiliation{
  Laser Physics Centre, RSPE, Australian National University, Canberra, ACT 2601, Australia
}

\author{C.~G. Poulton}
\affiliation{
 School of Mathematical and Physical Sciences, University of Technology Sydney, NSW 2007, Australia \\
}

\author{B.~J. Eggleton}
\affiliation{
Institute of Photonics and Optical Science (IPOS), The University of Sydney Nano Institute, School of Physics, University of Sydney, Sydney, NSW, Australia
}


\begin{abstract}
Stimulated Brillouin scattering drives a coherent interaction between optical signals and acoustic phonons and can be used for storing optical information in acoustic waves. An important consideration arises when multiple optical frequencies are simultaneously employed in the Brillouin process: in this case the acoustic phonons that are addressed by each optical wavelength can be separated by frequencies far smaller than the acoustic phonon linewidth, potentially leading to crosstalk between the optical modes. 
Here we extend the concept of Brillouin-based light storage to multiple wavelength channels. We experimentally and theoretically show that the accumulated phase mismatch over the length of the spatially extended phonons allows each optical wavelength channel to address a distinct phonon mode, ensuring negligible crosstalk and preserving the coherence, even if the phonons overlap in frequency. This phase-mismatch for broad-bandwidth pulses has far-reaching implications allowing dense wavelength multiplexing in Brillouin-based light storage, multi-frequency Brillouin sensing and lasing, parallel microwave processing and quantum photon-phonon interactions.
\end{abstract}

\maketitle



\section{Introduction}

Coherent interactions between the optical and the acoustic domain enable breakthrough functionalities in fields such as light storage~\cite{Zhu2007, Fiore2011,Merklein2017}, non-reciprocal systems~\cite{Dong2015,Ruesink2016,Sohn2018}, generating macroscopic quantum  states~\cite{Verhagen2012,Galland2014} and microwave photonics~\cite{Balram2015,Fang2016a,Merklein2016a}, with a number of recent developments in integrated photonic platforms~\cite{Eggleton2013,VanLaer2015,Kittlaus2015}. Within this broader field, interactions between optical modes and traveling acoustic waves via stimulated Brillouin scattering (SBS) are of particular interest because coupling occurs in the absence of a cavity; the optical field can therefore couple to a continuum of phonon modes rather than being limited to a discrete set of acoustic modes. A consequence of this coupling to the continuous field is the possibility of exciting multiple, closely-spaced acoustic modes. This feature is important for wavelength multiplexed light processing, particularly light storage, in which information can be transferred back and forth between the optical and acoustic domains over a continuous bandwidth limited only by the transparency window of the photonic waveguide ~\cite{Zhu2007,Merklein2017}. Because the Brillouin-based information transfer can be operated over a large frequency region, the important question arises if multiple optical signals at different frequencies can be simultaneously transferred. 

\begin{figure}[!b]
   \centering
   \includegraphics[scale=0.54]{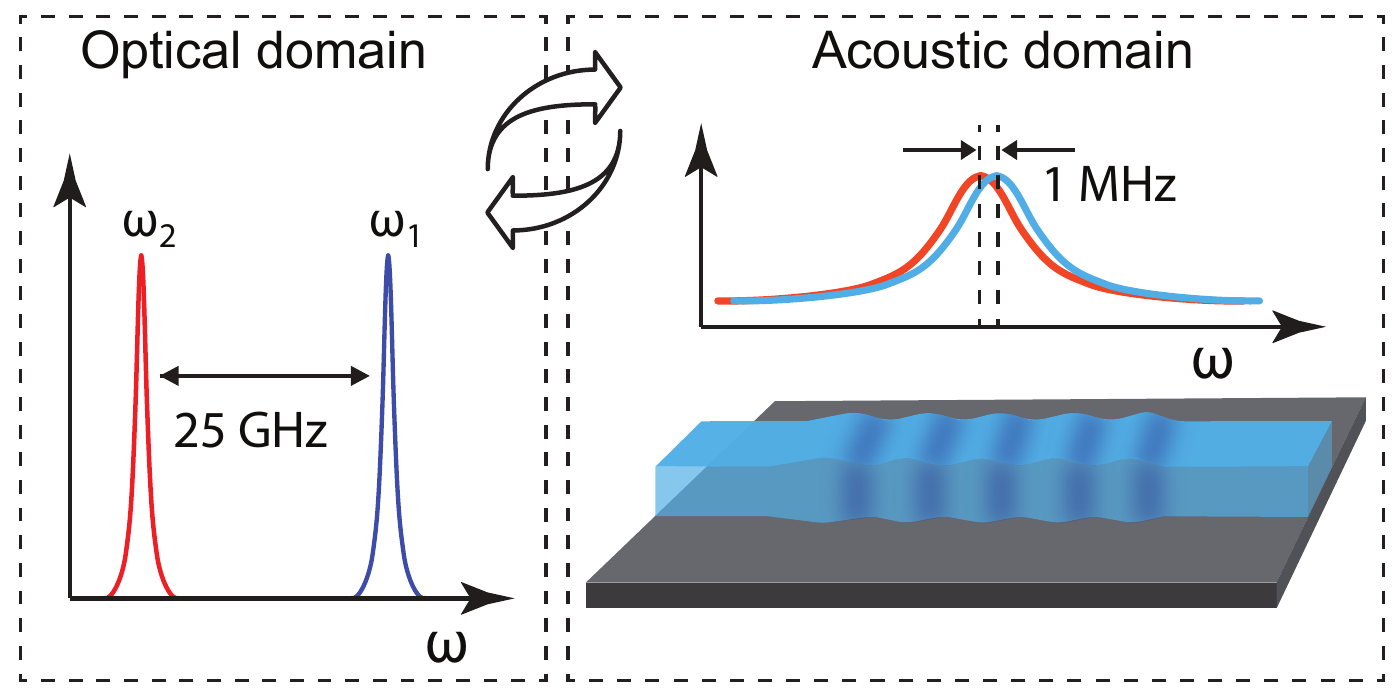}
 \caption{Coherent transfer between the optical and acoustic domain. The two optical wavelengths are separated by $2\pi\Delta F$, here about 25\,GHz. They are converted to two acoustic phonons which are separated by about 1\,MHz, which lays well within the Brillouin linewidth.}
 \label{fig0}
\end{figure}

The potential difficulty in transferring multiple optical signals becomes clear when considering that for optical frequencies spaced by hundreds of GHz, the spectral distance between the resulting acoustic lines can lie far below the acoustic linewidth (Fig. 1). One might therefore expect that this will result in significant crosstalk between simultaneously-written optical signals, which would put severe limitations on the applicability of Brillouin processes to multi-frequency signal processing. On the other hand, it is known from continuous-wave (CW) studies of multi-wavelength optical pumps on Brillouin processes ~\cite{Narum1986,Lichtman1987,Aoki1988} that the three-wave interaction between the optical pump wave, the Stokes wave and the acoustic wave prevents coupling to additional optical waves that are phase-mismatched. How these phase and frequency effects carry across to the more complicated situation of optical information storage, in which short pulses at multiple wavelengths are stored coherently in the acoustic domain, is the central question that we seek to address here. Previous experimental results for two optical wavelengths indicated that a simultaneous storage process is possible~\cite{Merklein2017}, however a thorough investigation of the crosstalk, the coherence of the process and the theoretical description has yet to be shown.

In this work, we demonstrate how optical pulses at different optical wavelengths can be simultaneously coherently stored in distinct spatially and temporarily overlapping travelling acoustic waves with negligible crosstalk. We use the feature of converting optical pulses to acoustic phonons and back in a Brillouin-based light storage configuration to probe individual acoustic phonon modes. This allows us to experimentally investigate the simultaneous excitation of several acoustic phonons and their respective cross-coupling. More specifically, we experimentally demonstrate a comprehensive study of the simultaneous storage of optical information at different wavelength channels, separated by 25\,GHz to 100\,GHz, as acoustic phonons. In the case of 25\,GHz, the acoustic phonons are separated in acoustic frequency as close as 1\,MHz (Fig. 1) and still no crosstalk is observed. We show that the stored information in these acoustic phonons can be unambiguously retrieved for each optical wavelength and that no measurable crosstalk to the upper or lower optical wavelength channel can be observed. We also demonstrate that the coherence of the Brillouin-based storage is not affected by parallel wavelength channels. A theoretical derivation for this phenomenon is given where we show that the phase accumulation along the pulse interaction length dampens the crosstalk between optical modes and mismatched acoustic waves. We theoretically study crosstalk depending on the pulse bandwidth, the pulse shape and the spacing of the different wavelength channels. This work therefore conclusively demonstrates that the Brillouin storage involves distinct phonons, notwithstanding their overlap in both the frequency and spatial domains.

\section{Multi-wavelength Brillouin transfer}

\begin{figure}[!b]
   \centering
   \includegraphics[scale=0.29]{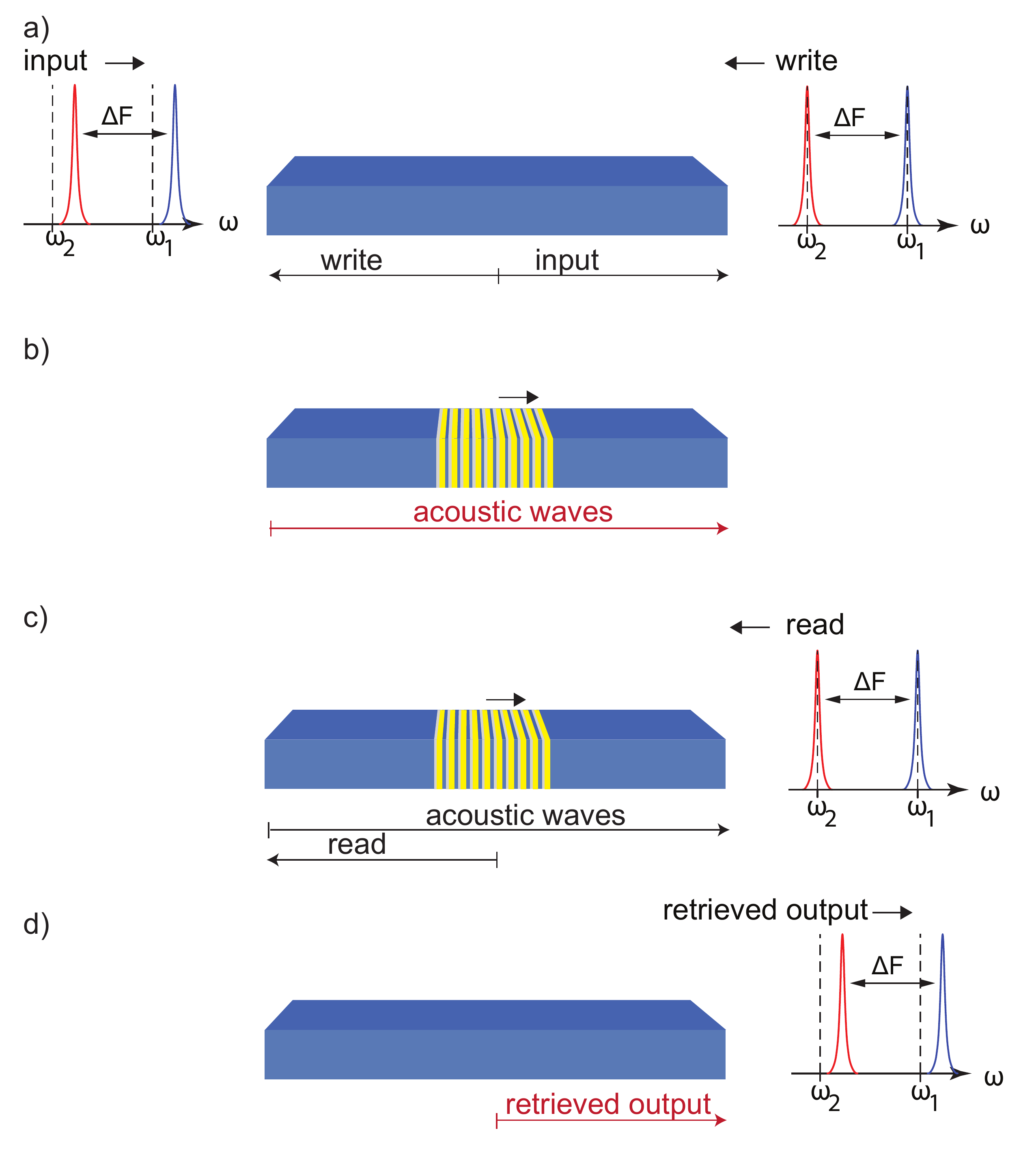}
 \caption{SBS-based coherent transfer principle: a) A data pulse and a counter-propagating write pulse enter the photonic chip for each optical wavelength (separated by $\Delta F = $ 25\,GHz); the data pulses are up-shifted in frequency by the respective Brillouin frequency shift.  b) The data pulses are depleted and two acoustic waves at close acoustic frequencies ($\Delta F = $1\,MHz) are created via SBS. c) For each optical wavelength, a read pulse enters the waveguide and depletes the acoustic waves by the reverse process. d) The retrieved data pulses, which are a delayed version of the original pulses, then exit the photonic chip, unambiguously separated by $\Delta F$.}
 \label{fig1}
\end{figure}

The detailed principle for the coherent tranfer for multiple wavelengths is depicted in Fig.~\ref{fig1}. In this scheme, an optical data pulse at frequency $\omega_{\mathrm{data}}$
interacts with two counter-propagating control pulses ``write'' and ``read'' at $\omega_{\mathrm{control}} = \omega_{\mathrm{data}} - \Omega$, where $\Omega$ is the Brillouin frequency shift (BFS) of the waveguide. The optical data pulse is transferred to the acoustic domain by a ``write'' pulse and transferred back to the optical domain by a ``read'' pulse. For each wavelength channel one optical data pulse enters the waveguide from one side (Fig.~\ref{fig1}a). From the other side, one ``write'' pulse for each wavelength counter-propagates the data pulses and completely depletes them. The wavelengths of the control pulses are 1550.05\,nm (wavelength 1) and 1550.25\,nm (wavelength 2). The data pulses are frequency up-shifted by the respective BFS which is around 7.8\,GHz in chalcogenide (As$_2$S$_3$). Two travelling acoustic waves at frequencies $\Omega_1$ and $\Omega_2$ are written into the waveguide (Fig.~\ref{fig1}b). The acoustic waves are then depleted by a ``read'' pulse at each wavelength channel (Fig.~\ref{fig1}c). This read-out process converts the acoustic waves back to the optical domain. We show that both data pulses at the respective wavelength are retrieved unambiguously (Fig.~\ref{fig1}d). This process can be applied over the transparency window of the waveguide. 

The BFS for a single optical pulse at wavelength $\lambda_{\mathrm{data}}$ is given by~\cite{Boyd2003}:
\begin{equation}
\Omega=\frac{2 n_{\mathrm{eff}}V_{A}}{\lambda_{\mathrm{data}}}~,
\label{BFS_eq}
\end{equation}
where $V_{A}$ and $n_{\mathrm{eff}}$ are the acoustic velocity and effective refractive index in the waveguide, respectively.
Multiple optical pulses with closely-spaced frequencies will result in multiple acoustic waves, 
each with frequency given by (\ref{BFS_eq}), however the absolute difference between any two acoustic frequencies $\Delta \Omega$ can be quite small due to the flat acoustic dispersion: for a chalcogenide rib waveguide carrying two optical channels separated by 25\,GHz at 1550.05\,nm and 1550.25\,nm, we obtain from  Eq. (\ref{BFS_eq}) a difference in acoustic frequencies $\Delta \Omega \sim 4$\,MHz. This is far narrower than the acoustic linewidth, which is around $30$\,MHz in this platform; it might therefore be expected that, if both optical  pulses are written to the same part of the waveguide simultaneously, a single acoustic mode would be generated with a spectral width broadened by $\Delta \Omega$. This would result in significant crosstalk  between the two optical signals, and put strict limits on the spectral density of optical signals that can be handled by the Brillouin transfer process. 

\begin{figure}[!t]
   \centering
   \includegraphics[scale=0.5]{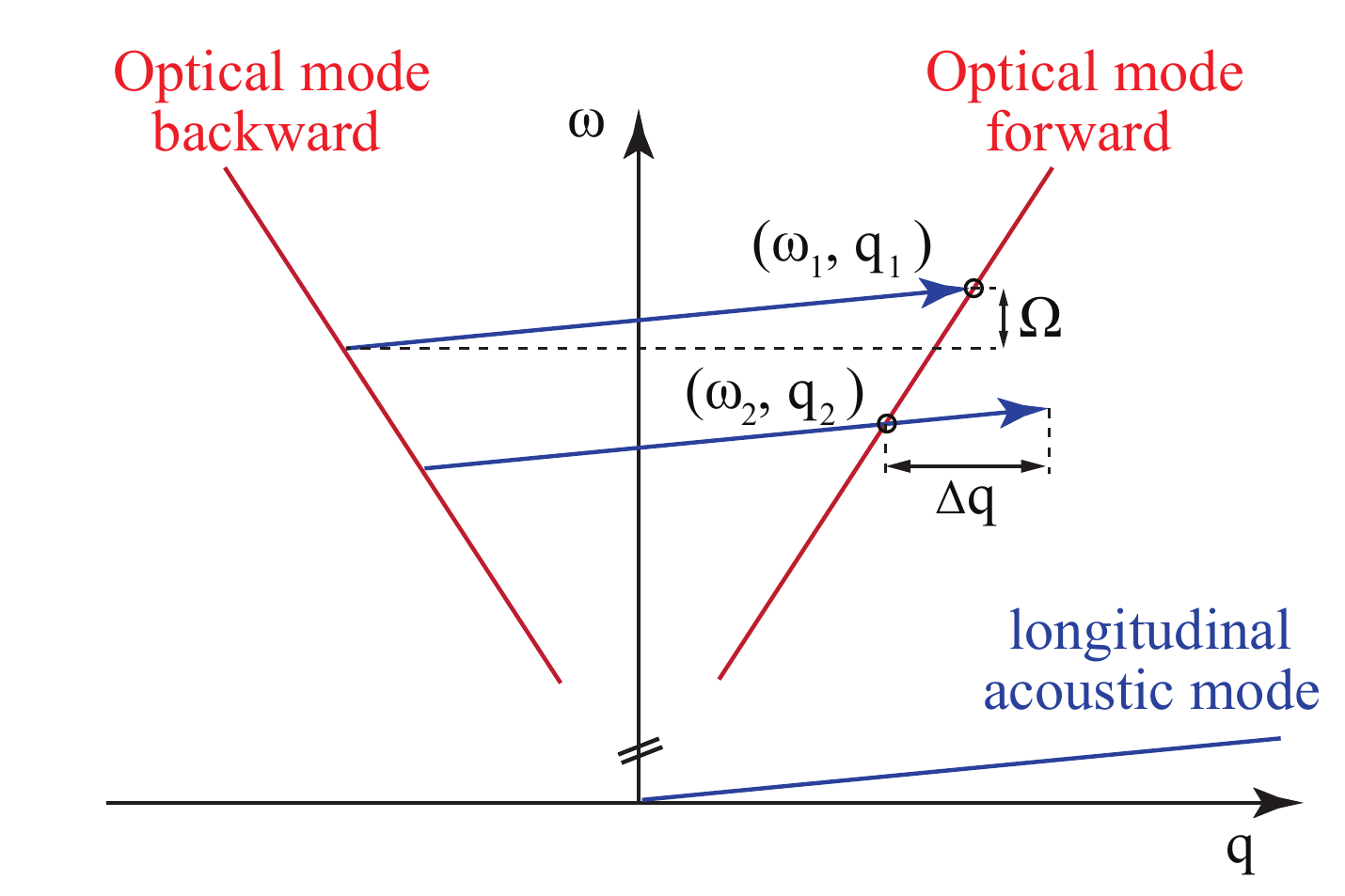}
 \caption{Dispersion relation for the backward Brillouin scattering process at two different optical wavelengths.}
 \label{figDisp}
\end{figure}

However, acoustic waves generated from closely-spaced data pulses can differ significantly in wavenumber accummulated over the long length of the interaction \cite{Narum1986,Lichtman1987,Aoki1988}. The dispersion relation for two backward Brillouin processes is shown in Fig.~\ref{figDisp}. For two optical pulses separated by $\Delta F$, the difference in  wavenumber between the acoustic modes is given by
\begin{equation}
\Delta q = 4 \pi n_{\mathrm{eff}} \Delta F / c.
\label{DeltaqEq}
\end{equation} 
 
For an optical spacing of 25\,GHz and an effective index of 2.5, this corresponds to a wavenumber mismatch of $\Delta q \sim 0.1 \times 10^4$m$^{-1}$.
Any interaction length that is longer than approx. 200\,$\upmu $m will then result in phase accumulation that dampens the interaction between the  control pulse and the unmatched acoustic wave. We examine the formalism of the phase mismatch further in section IV.

\begin{figure*}
   \centering
   \includegraphics[scale=0.5]{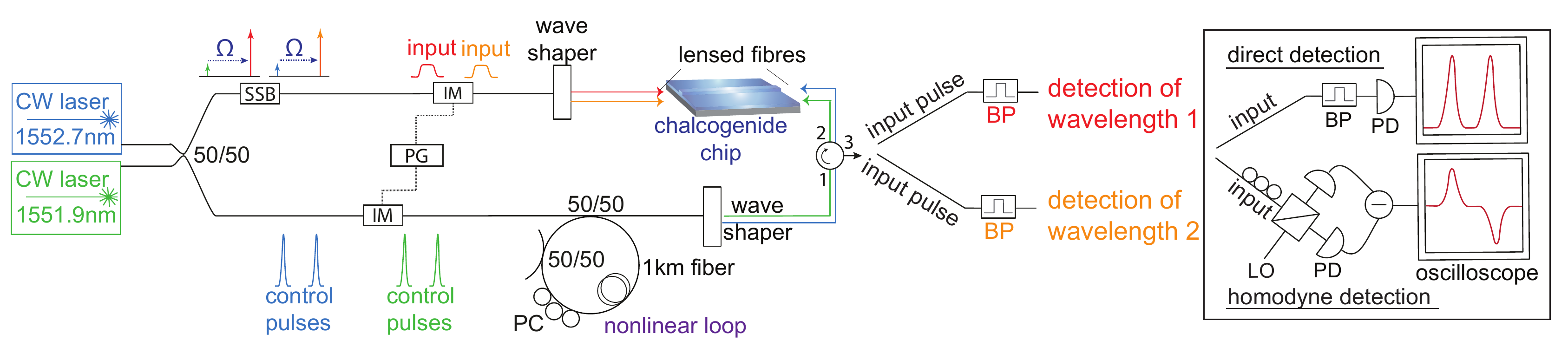}
 \caption{Experimental setup for studying the coherent multi-wavelength Brillouin interaction: PC polarization controller, SSB single-sideband modulator, IM intensity modulator, PG pulse generator, BP bandpass filter, LO local oscillator, PD photo diode.}
 \label{fig3}
\end{figure*}

\section{Experimental results}

For the experimental studies, we use a highly nonlinear As$_2$S$_3$ rib waveguide~\cite{Madden2007}, arranged as a 22cm-long spiral on a photonic chip. The experimental setup is depicted in Fig.~\ref{fig3}. Two continous wave (CW) distributed feedback lasers that are separated by 25\,GHz, are used for the generation of two optical signal streams at a different wavelength. Both wavelength channels are combined in a coupler and then divided into a data arm and an control pulse arm. 

In the data arm, the signals at both wavelengths pass through a single-sideband modulator, in order to frequency up-shift the lasers by the BFS (single-sideband operation and carrier suppression). The CW light is then carved into 2ns-long pulses (repetition rate 500\,kHz). The same single-sideband modulator can be used for both wavelengths, as the change in BFS is only 1\,MHz, which is more than two orders of magnitude less than the  bandwidth of the used data pulses. The wavelengths are then separated by a multi-channel Finisar WaveShaper which switches the data streams on and off at the specific wavelength. The approximate data pulse peak power is 100\,mW.

On the control pulses arm, a write and a read pulse (2\,ns long, repetition rate 500\,kHz, 3\,ns time delay between the write and read pulse) is carved into the CW light of the laser and a second Finisar WaveShaper is used to select and switch the respective wavelength channel. A nonlinear loop ensures that detrimental noise stemming from a limited extinction ratio of the modulator and amplified spontaneous emission of the amplifier is filtered and only high-amplitude coherent pulses are transmitted. The approximate control pulse peak power is 10\,W. For each channel, a 3GHz-narrow filter is used in front of an amplified photodiode to directly detect the data pulses. On wavelength channel 1, a data pulse at $\omega_{\mathrm{data 1}}$ is transferred by corresponding control pulses $\omega_{\mathrm{control 1}}$ into an acoustic wave $\Omega_{1}$. Simultaneously, another data pulse at $\omega_{\mathrm{ 2}}$ - 25\,GHz apart from data 1 - is transferred into a second acoustic wave $\Omega_{2}$ by respective control pulses $\omega_{\mathrm{control 2}}$ in channel 2. 

The experimental results are shown in Fig.~\ref{fig4a} to ~\ref{fig6}. The efficiency of the depletion of the initial data pulses when writing to the acoustic domain is more than 80 \% and the overall efficiency of the retrieved data is approximately 25 \%. The reason that the depletion is not 100 \% can be attributed to limited control pulse power and losses in the waveguide. The pulse shape is roughly conserved, however this can be still improved by using shorter control pulses with a broader bandwidth on the expense of higher power requirements.

We investigate possible crosstalk for two cases: firstly, we show that a data pulse on a second wavelength channel is not transferred to an acoustic wave (Fig.~\ref{fig4a}); secondly, we demonstrate that a pre-written acoustic wave cannot be read-out by control pulses on a second wavelength channel (Fig.~\ref{fig4b}). Then, we study simultaneous storage on both wavelengths (both data pulses and control pulse pairs enter the chip) with comparison to single-wavelength operation on wavelength 1 (Fig.~\ref{fig5}). We show these results for different frequency spacing of the two channels from 100\,GHz down to 25\,GHz (Fig.~\ref{figO}). Finally, we study the simultaneous storage of the optical phase and show that coherent storage at multiple wavelengths with negligible crosstalk is possible (Fig.~\ref{fig6}).

\subsection{Crosstalk study}

\begin{figure}
   \centering
   \includegraphics[scale=0.9]{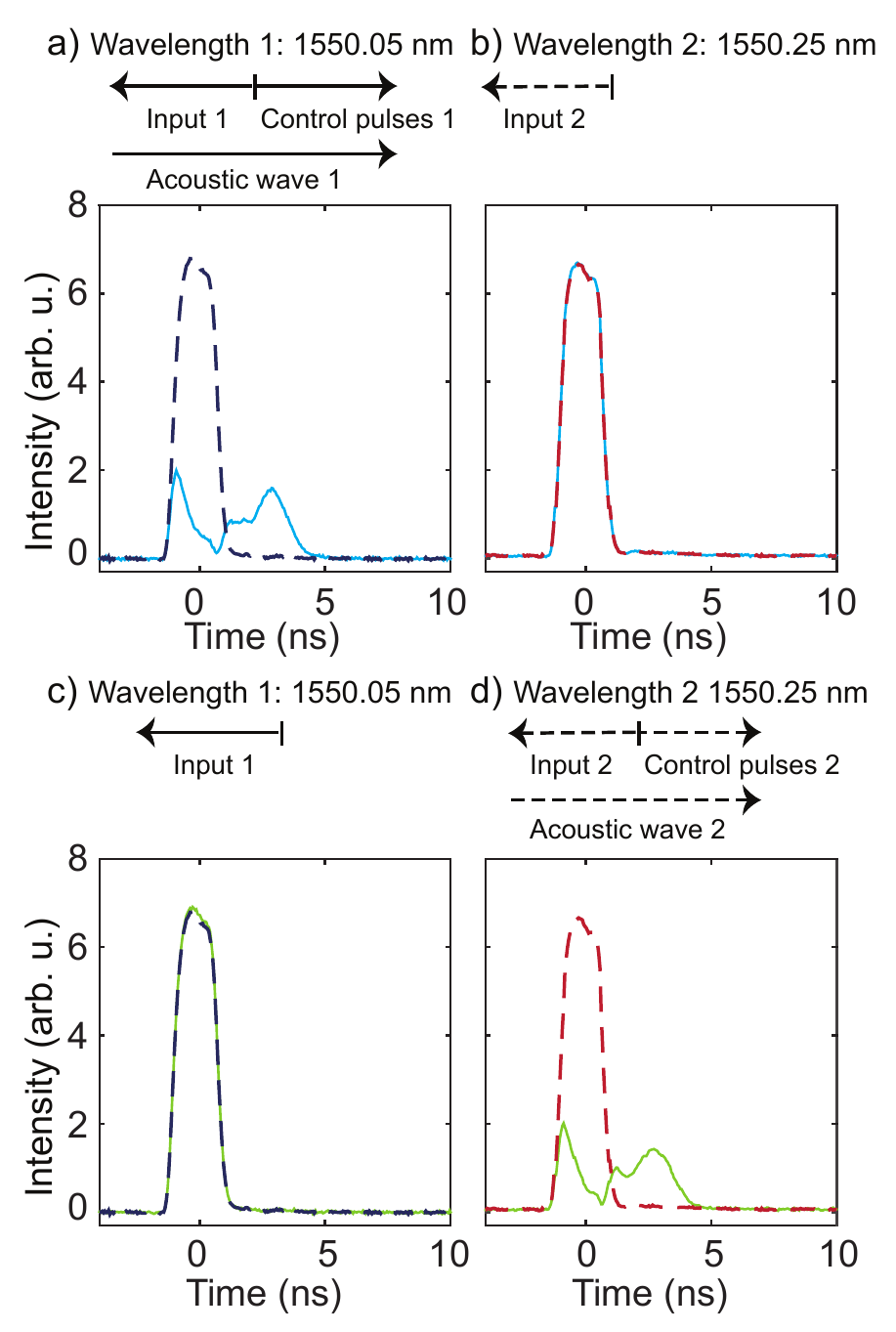}
 \caption{Wavelength 1 and 2 are separated by 25\,GHz; a) The original data pulse (dashed line) gets depleted and is retrieved 3\,ns later when the control pulses are switched on (light blue solid line); therefore, the coherent transfer to the acoustic domain is operating at wavelength 1. b) The data pulse at wavelength 2 is not affected, when the coherent transfer is operating at wavelength 1 (light blue line); the red dashed line represents the data pulse when no coherent transfer is taking place at wavelength 1. c) and d) Coherent transfer at wavelength 2 while the data at wavelength 1 is not affected by it.}
 \label{fig4a}
 \end{figure} 

We study possible crosstalk by demonstrating that the transfer and retrieval processes do not interfere between different optical wavelengths. Therefore, we consider two cases:

(1) the transmission at one wavelength is not affected by the opto-acoustic interaction at a second wavelength;

(2) the control pulses at one wavelength cannot interact with an acoustic wave generated by the optical pulses at the other wavelength.

In both cases we show that there is negligible crosstalk to the higher or lower wavelengths.

The first case is shown in Fig.~\ref{fig4a}. We consider that the coherent transfer is operated at wavelength 1 (data 1 and control pulses 1 are switched on, Fig.~\ref{fig4a}a) while control pulses 2 are switched off at wavelength 2 and hence data 2 is transmitted unaffectedly (Fig.~\ref{fig4a}b, light blue graph). The dashed red graph in Fig.~\ref{fig4a}b corresponds to the transmitted data pulse 2 when the pulses are switched off at wavelength 1. There is no visible difference in the transmitted data pulse at wavelength 2, if the pulses at wavelength 1 are switched on or off. Therefore, there is no measurable coupling into a lower wavelength channel. We also investigate possible coupling to the higher wavelength channel, by swapping the role of channel 1 and 2 (Fig.~\ref{fig4a}c and d). The coherent transfer is operated at channel 2 while a data pulse passes unaffected through channel 1. Both measurements show that no crosstalk to the higher or lower wavelength channel takes place. It means that the control pulses do only transfer the information of the optical pulse to the acoustic domain at the specific wavelength channel.

\begin{figure}[!t]
   \centering
   \includegraphics[scale=0.9]{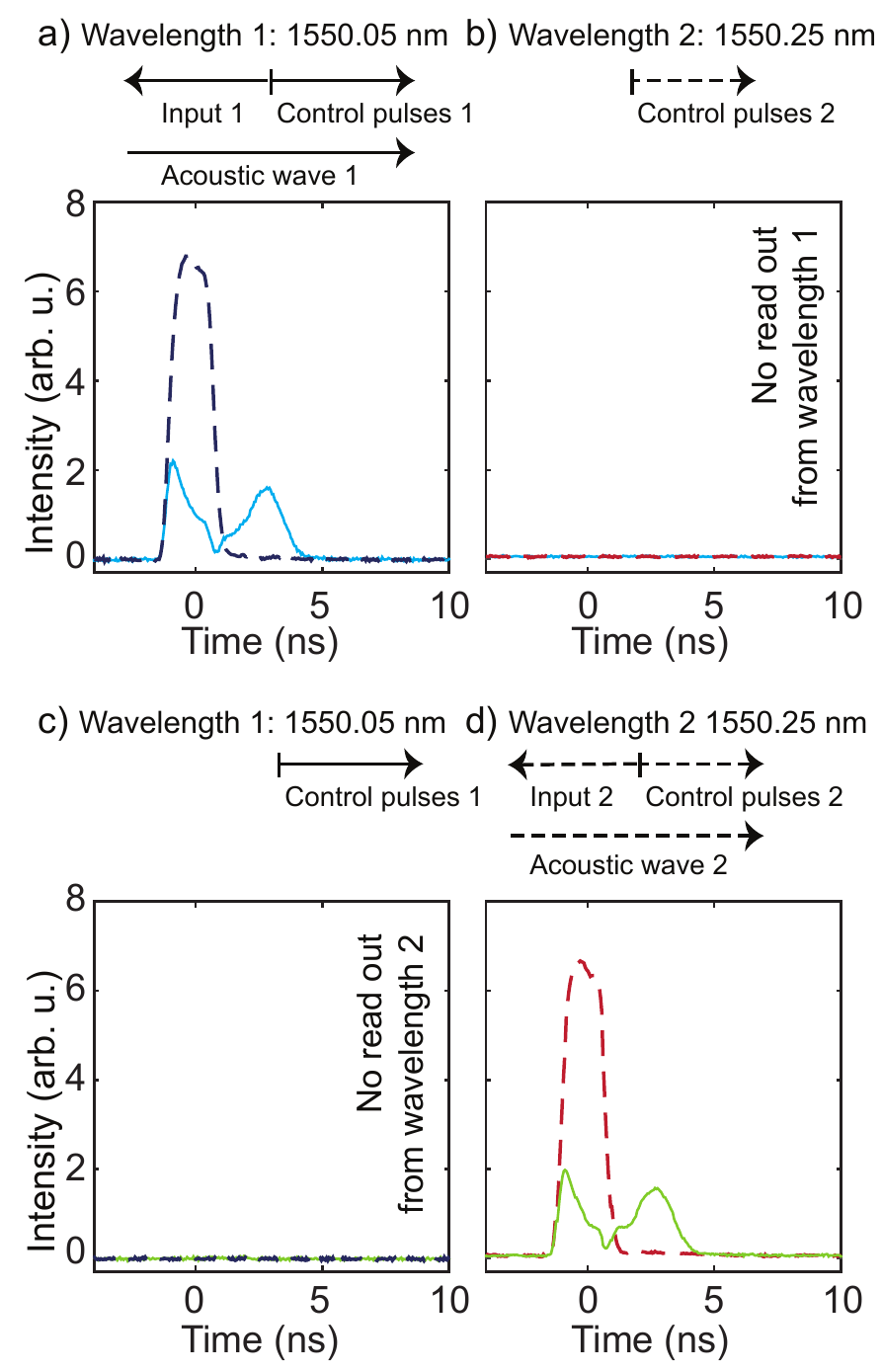}
 \caption{a) The coherent transfer to the acoustic domain and back is operating at wavelength 1; b) no read-out is observed at wavelength 2 (light blue solid line); the dashed line represents the reference noise level of the photo diode as comparison. c) and d) The coherent transfer at wavelength 2 is operating while no read-out is observed at wavelength 1 (green line); the dark blue dashed line is the reference noise level of the photodiode.}
 \label{fig4b}
 \end{figure}

The second case is demonstrated in Fig.~\ref{fig4b}. We consider that the coherent transfer is operated on wavelength channel 1 while no data pulse is transmitted at channel 2 but the control pulses $\omega_{\mathrm{control 2}}$ are operating on wavelength channel 2 (Fig.~\ref{fig4b}a,b). It shows that the control pulses $\omega_{\mathrm{control 2}}$ do not couple to the acoustic wave $\Omega_{\mathrm{1}}$ that has been created at wavelength channel 1 and therefore no delayed read-out is observed. In Fig.~\ref{fig4b}c and d the roles of the wavelength channels are changed and the coherent transfer is operated at channel 2 while no read-out pulse is observed on channel 1. This indicates that the control pulses cannot read out an acoustic phonon at a separate wavelength channel. All of the measurements have been carried out by simultaneously observing two channels on the oscilloscope.

\subsection{Simultaneous coherent opto-acoustic transfer on both channels}

After having shown that no measurable crosstalk between two wavelength channels, separated by 25\,GHz, is observed, we investigate the simultaneous coherent transfer on both channels, in order to prove that the acoustic phonons, spatially and temporally overlapping, do not interact with each other.

\begin{figure}[!t]
   \centering
   \includegraphics[scale=0.9]{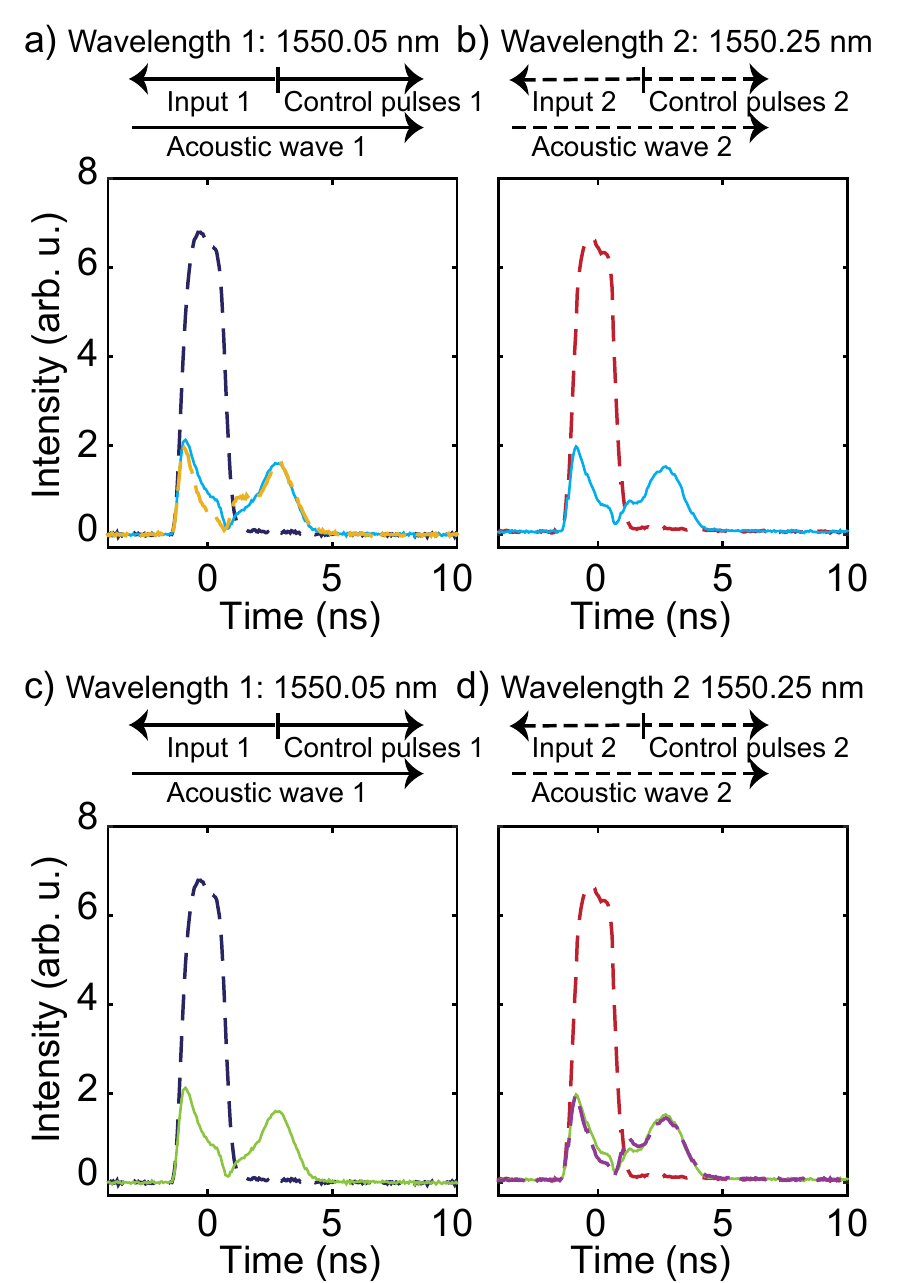}
 \caption{a) and b) Simultaneous operation of the coherent transfer at wavelength 1 and 2. The dashed lines represent the original data pulses. The original data pulses are depleted and retrieved after 3\,ns at both wavelengths (solid lines). The yellow graph in a) refers to a coherent transfer at wavelength 1 without memory operation at wavelength 2. c) and d) Comparison of simultaneous 2-channel storage operation to single-operation at wavelength 2 (violet solid line).}
 \label{fig5}
 \end{figure}

The transmitted data pulses 1 and 2 are shown in Figs.~\ref{fig5}a and b (dashed lines). They are both directly detected on two different photodiodes after being filtered out with two narrowband filters. When sending in the control pulses on both channels, the data pulses get transferred to the acoustic domain, displayed at position 0\,ns, and are retrieved 3\,ns later (Fig.~\ref{fig5}a,b, solid lines). To confirm that there is no interaction between the two acoustic phonons created in the two channels, we switch off channel 2 and record the coherent transfer on channel 1 which corresponds to the yellow graph in Fig.~\ref{fig5}a. We see that the transfered and retrieved data pulses overlap perfectly which means that the parallel operation at channel 2 does not influence the operation at channel 1. The same scenario has been tested for the higher wavelength channel (Fig.~\ref{fig5}b,c).

\begin{figure}[!t]
   \centering
   \includegraphics[scale=0.7]{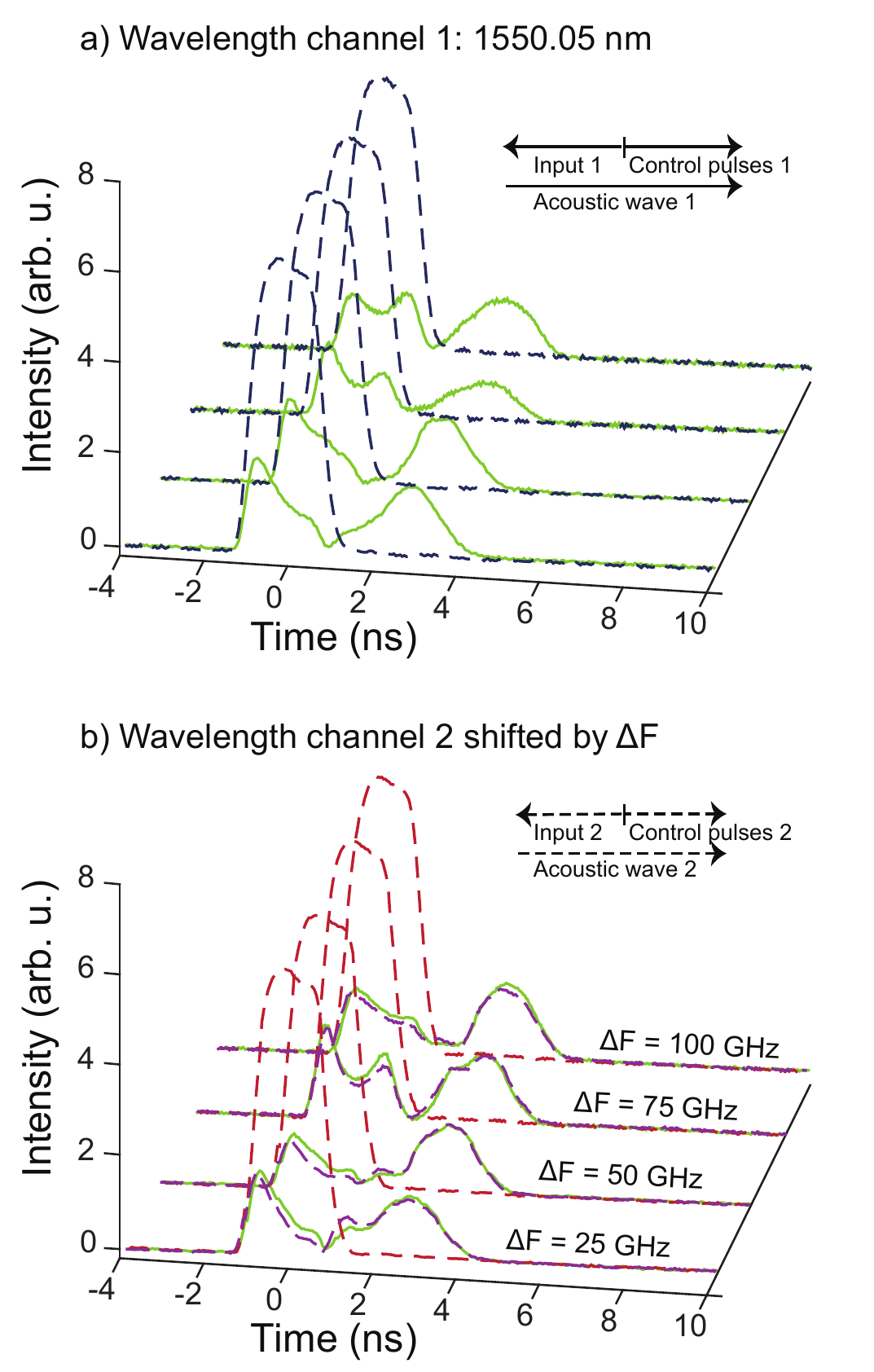}
 \caption{Simultaneous storage at two wavelength channels for different frequency spacing of the two channels: 25\,GHz, 50\,GHz, 75\,GHz and 100\,GHz. The dashed lines correspond to the undepleted data pulses and the solid lines to the depleted and retrieved data. 
 a) Wavelength channel 1. b) Wavelength channel 2. The violet solid line corresponds to single-channel storage at wavelength channel 2 only as a comparison to parallel storage operation at both wavelengths.}

 \label{figO}
 \end{figure}

All previous measurement have been carried out for a frequency spacing of 25\,GHz. We also swept the frequency spacing in 25GHz-steps up to 100\,GHz and performed the crosstalk study (both cases) as well as simultaneous storage on both channels. The experimental results for simultaneous storage in comparison to storage operation only on wavelength 2 are demonstrated in Fig.~\ref{figO}. It can be deduced that no obvious difference in single-channel operation and parallel storage is observed. Both cases of the crosstalk study, which is not visualized here, also show that no crosstalk between the two channels is observed at any of the chosen frequency spacing values.

\subsection{Phase retrieval in parallel wavelength channels}
\begin{figure}[!t]
   \centering
   \includegraphics[scale=0.71]{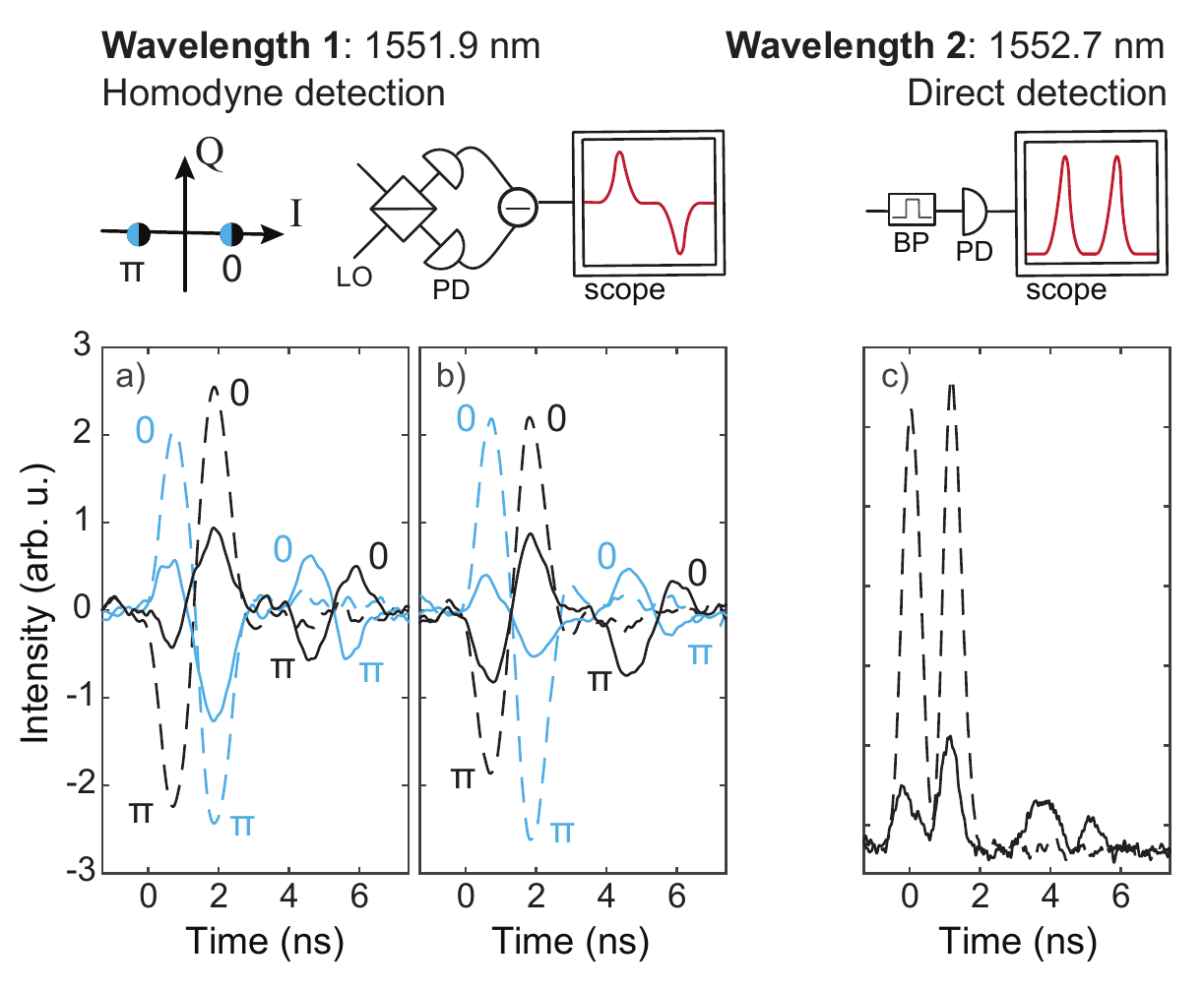}
 \caption{Storage of the optical phase at wavelength 1 and measurement of the stored amplitude at wavelength 2. The dashed lines correspond to the undepleted data pulses and the solid lines to the depleted and retrieved data. a) and b) Homodyne detection at wavelength 1 of two pulses that were encoded with two different optical phases 0 and $\pi$ (depicted in the constellation diagram in the inset). The optical phase is detected by interference with a local oscillator at the same wavelength. a) Operation of the coherent transfer at wavelength 1 only (without operation at wavelength channel 2). b) Storage at wavelength channel 1 (while simultaneously data is stored at wavelength 2). c) Simultaneously to wavelength 1: direct detection of two pulses at wavelength 2, 100\,GHz apart from wavelength 1.}
 \label{fig6}
 \end{figure}

As SBS is a coherent process, the optical phase can be transferred and retrieved from the acoustic wave \cite{Merklein2017}. In this section, we experimentally show coherent storage at multiple wavelength channels. As the optical phase is generally more sensitive to crosstalk and the crosstalk relies on a phase mismatch that is accumulated over the spatial extension of the phonon, proving the retrieval of the optical phase is an essential part for coherent multi-wavelength opto-acoustic storage. The experimental results here show the retrieval of the optical phase on one wavelength channel, while a data pulse is stored simultaneously at wavelength channel 2, separated by 100\,GHz. The optical phase has been encoded by adjusting the bias on the intensity modulator in the data arm. The detection of the phase is done by interfering the data pulses with a CW local oscillator (approximately 1\,mW) in a homodyne detection scheme (Fig.~\ref{fig3}). As comparison, we first show the phase retrieval at single-channel operation (Fig.~\ref{fig6}a, channel 1). Two pulses are encoded with a relative phase of $\pi$ (dashed graph in Fig.~\ref{fig6}a). When switching on the control pulses, the data pulses get depleted and are retrieved 4\,ns later (solid graph in Fig.~\ref{fig6}a). The optical phase is retrieved as $0$ and $\pi$ or $\pi$ and $0$, respectively. Afterwards, we study the storage of the optical phase at wavelength channel 1, while operating a second storage process at wavelength channel 2 (Figs.~\ref{fig6}b and c). As can be seen in Fig.~\ref{fig6}b, the operation at channel 1 remains unaffected while simultaneously storing light at wavelength channel 2 (Fig.~\ref{fig6}c). The light storage at wavelength channel 2 is only measured in direct detection due to equipment restrictions. Demonstrating the ability of a multi-wavelength coherent opto-acoustic transfer is important for delaying signals selectively in a coherent communication system and might be relevant for future quantum communication systems with enhanced capacities due to the use of multiple wavelengths.

It is not sufficient to show simultaneous multi-wavelength storage in terms of the amplitude of the data but also in terms of the optical phase, because the latter might be more sensitive to crosstalk between the channels. 
Moreover, the negligible crosstalk relies on a phase mismatch that is accumulated over the spatial extension of the phonon. Therefore, showing the coherent retrieval of the information - also at multiple wavelength channels - is essential.

 \section{Theory of Brillouin-based light storage for phase-mismatched pulses}

\begin{figure}[!t]
   \centering
   \includegraphics[scale=0.5]{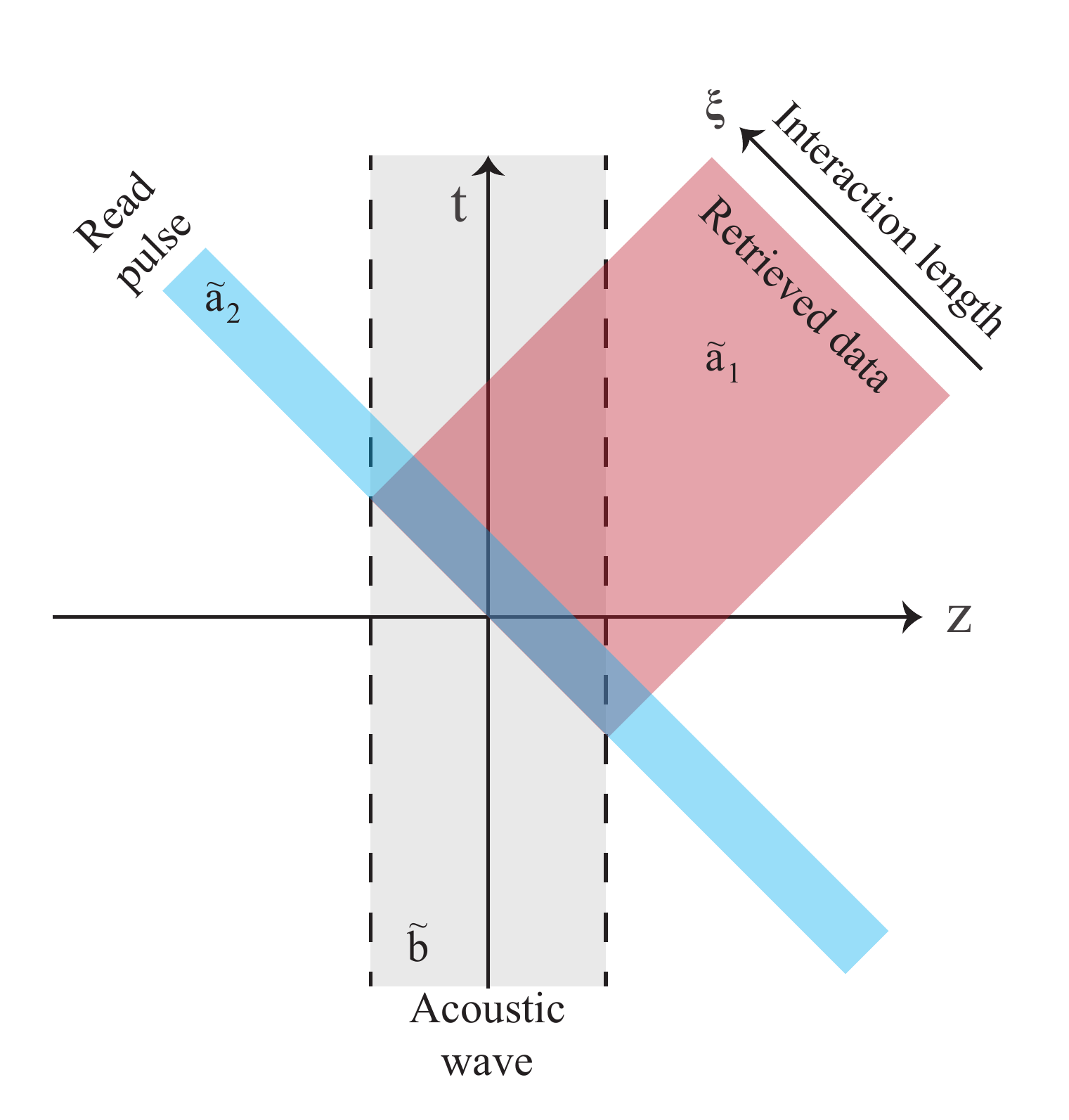}
 \caption{Conversion of the acoustic phonon to a retrieved data pulse via the read pulse in the reference frame of the acoustic wave.}
 \label{ConvFig}
\end{figure}

The theory of acousto-optic light storage in the situation where the data pulses have multiple wavelengths can be obtained
by adapting the three-wave interaction model of Dodin and Fisch \cite{dodinfisch}. We consider the situation where
an acoustic wave has the potential to couple to two optical data pulses: one which is perfectly phase-matched
to the acoustic mode, while the second is offset by a frequency $\Delta F$ and so is phase-mismatched by $\Delta q$ (Eq. (\ref{DeltaqEq})).
The proportion of acoustic power transferred to the mismatched term is a measure of the crosstalk between the adjacent frequencies.
The three-wave interaction model allows the quantitative analysis of two different forms of crosstalk - namely how much of the data optical power is transferred to an acoustic wave by a control pulse with a phase-mismatch  $\Delta q$ and secondly how much power of an excited acoustic mode is read out by a control pulse with a phase-mismatch $\Delta q$. In the following we will focus on the second case, and note that the rate of transfer is in fact the same for both processes if the read/write pulses are identical.  
To compute the crosstalk between adjacent frequencies, we consider a data pulse with mode amplitude $a_1(z,t)$, a control pulse $a_2(z,t)$ and an acoustic wave $b(z,t)$. The coupled mode equations that relate the mode amplitudes are given by \cite{Wolff2015}

\begin{subequations}
\label{coupled1}
\begin{align}
 \label{eq:a}
\left(\partial_z + \frac{1}{\vop} \partial_t\right) a_1 & =-{\rm i} \frac{\omega Q}{\Pop} a_2 b^* 
e^{{\rm i} \phi(z,t)} \\ 
 \label{eq:b}
\left(\partial_z - \frac{1}{\vop} \partial_t\right) a_2 & = {\rm i} \frac{\omega Q^*}{\Pop} a_1 b 
e^{-{\rm i} \phi(z,t)} \\
 \label{eq:c}
\left(\partial_z + \frac{1}{\vac} \partial_t + \frac{1}{\vac} \alpha \right) b & = -{\rm i} \frac{\Omega Q}{\Pac} a_1^* a_2 
e^{{\rm i} \phi(z,t)}
\end{align}
\end{subequations}

\noindent where the effect of the phase-mismatch is incorporated via the term $\phi(z,t) = \Delta q z$. In Eqs. (\ref{coupled1}) the
quantity $Q$ is the SBS coupling coefficient given in \cite{Wolff2015}, $\Pop$ and $\Pac$ are the modal powers, $\omega$ and $\Omega$ are the frequencies, \(\alpha\) is the acoustic damping rate and $\vop$ and $\vac$ are the group velocities of the optical and acoustic modes, respectively. 
A schematic of the propagation of the optical data pulse and the control (read) pulse is shown in Fig.~\ref{ConvFig}. Here we have assumed that the control pulse propagates in the negative $z$ direction with a velocity $-\vop$ and that all optical modes share the same power normalisation. In the reference frame of the acoustic wave (Fig.~\ref{ConvFig}), the control read pulse depletes the acoustic wave over the interaction length while travelling through the localised acoustic wave. The data pulse is retrieved over the interaction length and propagates in the positive $z$ direction.

We assume that the frequency shift associated with the Brillouin process is sufficiently small that the frequencies of the control and data pulses can be taken to be identical, although their difference in phase is non-negligible. Following \cite{dodinfisch}, we move to a dimensionless coordinate system that co-propagates with the control pulse, such that 
\begin{equation}
\zeta = \left(z + \vop t \right)/Z_0 ~~~, \tau = t/T_0 ~,
\label{Z0Tau}
\end{equation}
where $Z_0$ and $T_0$ are constants that will be chosen to simplify the resulting system of equations. 
Before determining these parameters, we first make the approximation that the control pulse is both short 
enough in duration and sufficiently strong that the 
the interaction is ``static'' in the sense that all derivatives with respect to $\tau$ are small with respect to
the derivatives with respect to $\zeta$. This assumption holds provided 
$|a_2/L_r|>>|\omega Q a_1 b / P|$ for the peak amplitudes ~\cite{dodinfisch}, where $L_r$ is the pulse length of the read pulse. The static approximation states that the temporal variations of the acoustic and data pulses are small compared to the spatial derivatives in the moving reference frame of the write/read pulses. This will be the case for a sufficiently strong write/read pulse with a short pulse length, for which the entirety of the change in the data pulse and acoustic mode must occur over the length of the write pulse. We also assume that the effect of the interaction on the control pulse can be neglected.
Under these assumptions, and including the additional observation that
$\vac << \vop$, the equations (\ref{coupled1}) become
\begin{subequations}
\label{coupled2}
\begin{align}
\partial_\zeta a_1 &=-Z_0\frac{{\rm i} \omega Q}{2 \Pop} a_2 b^* e^{{\rm i} \phi(\zeta,\tau)} \\
\partial_\tau a_2 &=-T_0\frac{{\rm i} \omega Q^* \vop}{\Pop} a_1 b e^{-{\rm i} \phi(\zeta,\tau)} \\
\partial_\zeta b  &=-Z_0\frac{{\rm i} \Omega Q \vac}{\Pac \vop} a_2 a_1^* e^{{\rm i} \phi(\zeta,\tau)}.
\end{align}
\end{subequations}

We now define new amplitudes $\tilde{a}_1 = a_1 / C_1$ and $\tilde{b} = b^* / C_b$, and choose the scaling $C_1$ and $C_b$,
as well as $Z_0$ and $T_0$, such that the coupling coefficients in Eq. (\ref{coupled2}) are equal to unity;
this condition results in $Z_0 = \sqrt{2 \Pop \Pac v / \omega \Omega |Q^2| \vac}$ and $T_0 = Z_0 / |v|$. 
The resulting set of coupled equations for the data and control pulses is then
\begin{subequations}
\label{couplednorm}
\begin{align}
\partial_\zeta \tilde{a}_1 &= a_2 \tilde{b} e^{i \phi(\zeta,\tau)} \\
\partial_\zeta \tilde{b} &= -a_2^* \tilde{a}_1 e^{-i \phi(\zeta,\tau)}.
\end{align}
\end{subequations}

It is readily verified that these satisfy the Manley-Rowe equations $|\tilde{a}_1(0,\tau)|^2 + |\tilde{b}(0,\tau)|^2 = |N_0(\tau)|^2$. From this observation, and making use of the assumption that the shortness of the control pulse means that 
its phase is approximately constant across the interaction region, such that we can write $a_2(\zeta,\tau) = A_2(\zeta \tau) \exp(i \theta)$, where $A_2$ is real, the solution to (\ref{couplednorm}) takes the form:
\begin{subequations}
\label{sol1}
\begin{align}
\tilde{a}_1(\zeta,\tau) & = N_0(\tau) \cos \left( U(\zeta,\tau) + \varphi \right) \\
\tilde{b}(\zeta,\tau) & = -N_0(\tau) \sin \left( U(\zeta,\tau) + \varphi \right),
\end{align}
\end{subequations}
where $\varphi$ is the phase required to fit the initial conditions for the system given in Eq. (6): $\varphi = 0, \pi/2$ correspondig to the write and read processes, respectively, and
$N_0(\tau) = (|\tilde{a}_1(0,\tau)|^2 + |\tilde{b}(0,\tau)|^2)^{1/2}$.
For the case of phase-matched pulses, these solutions $\tilde{a}_1(\zeta,\tau)$ and $\tilde{b}(\zeta,\tau)$ are depicted in Fig.~\ref{TransferFig}a.
The function $U(\zeta, \tau)$ is an integral over the control pulse amplitude:
 \begin{figure}[!t]
   \centering
   \includegraphics[scale=0.4]{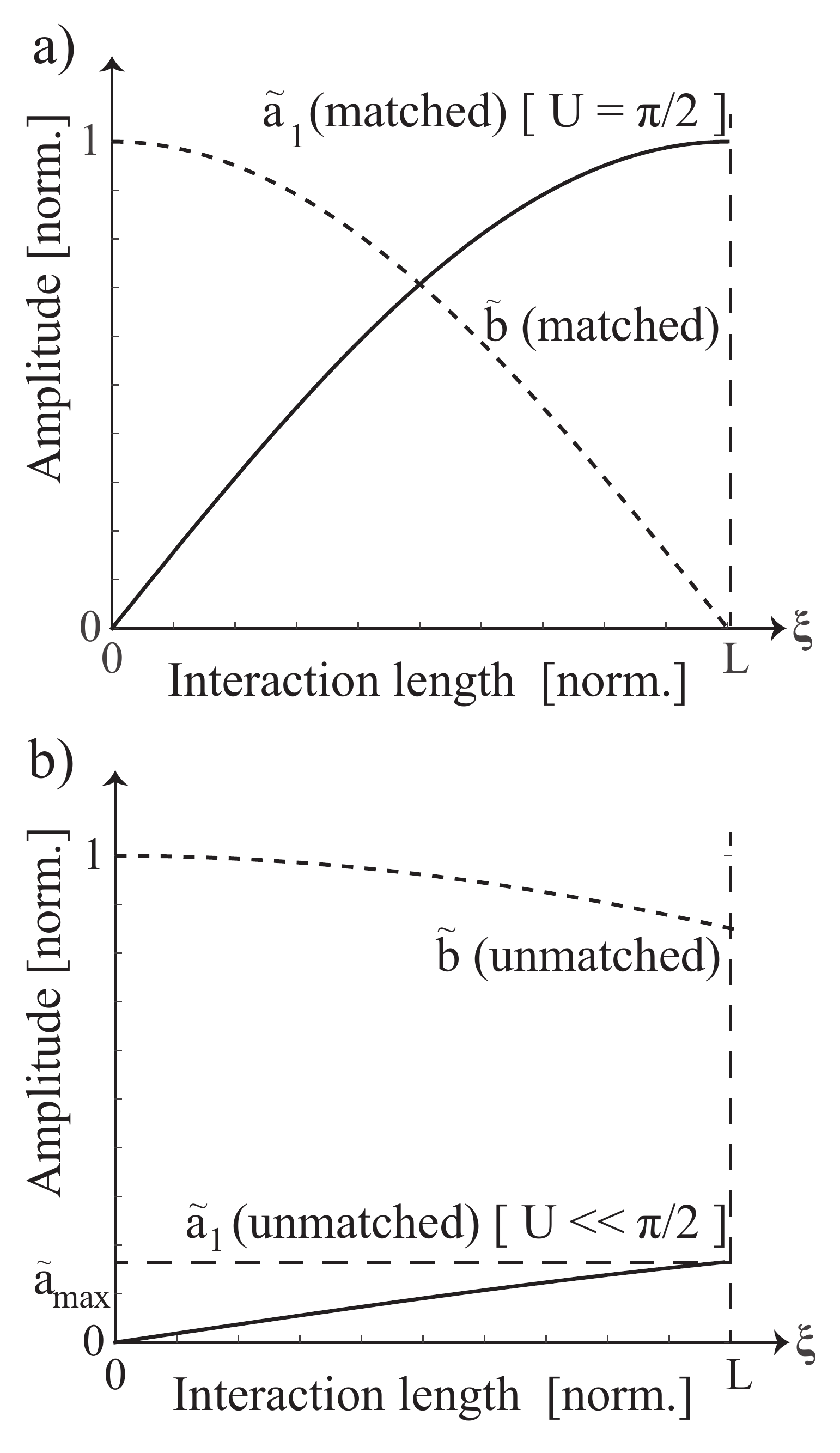}
 \caption{Amplitude of the transfer of the acoustic phonon to the data pulse for the case of a) phase-matched pulses (same wavelength channel), b) phase-mismatched pulse (other wavelength channel).}
 \label{TransferFig}
\end{figure}

\begin{equation}
U(\zeta,\tau) = 2 \int_{-\infty}^{\zeta} A_2(\zeta',\tau) e^{{\rm i} \phi(\zeta',\tau)} d\zeta'.
\label{Udef}
\end{equation}
The mode amplitude at the conclusion of a ``read'' interaction, with input acoustic amplitude $\tilde{b}_{\rm in}(0)$,
is then given by
\begin{equation}
\tilde{a}_1 (\zeta,\tau \rightarrow \infty) = \tilde{b}_{\rm in}(0) \sin\left(U_{\Delta q} \right) ~,
\end{equation}
where the total pulse area, written in the original coordinates $(z,t)$, is equal to 
\begin{equation}
U_{\Delta q}= \frac{1}{Z_0} \int_{-\infty}^{\infty} A_2(z,0) e^{{\rm i} \Delta q z} \rm{d}z ~.
\label{Udeltaq}
\end{equation}
with $U_0 := U_{\Delta q = 0}$ defined as the phase-matched pulse area. The power crosstalk is 
the ratio between the output powers of phase-matched and phase-mismatched data signals,
and is given by 
\begin{equation}
\chi = \frac{\sin^2\left(U_{\Delta q} \right)}{\sin^2\left(U_{0} \right)} ~.
\end{equation}

In the case where a phase-matched pulse would be completely converted we have $U_0 = \frac{\pi}{2}$, giving the crosstalk as
\begin{equation}
\chi = \sin^2\left(U_{\Delta q} \right).
\end{equation}

\noindent For the case of $U_{\Delta q} << \frac{\pi}{2}$, i.e. for phase-mismatched $\tilde{a}_1(\zeta,\tau)$ and $\tilde{b}(\zeta,\tau)$, the amplitude of the coherent transfer is shown in Fig.~\ref{TransferFig}b.

From Eq. (\ref{Udeltaq}) it can be seen that the effect of the phase-mismatch is to reduce the pulse area, ideally equal to
$\pi /2$, to some lower value. Because Eq. (\ref{Udeltaq}) is the Fourier transform of the mode amplitude, we conclude that the 
crosstalk will be reduced in direct relation to the spectral width of the data pulse: rapidly for Gaussian pulses and 
more slowly for square pulses. The crosstalk for both these pulses can be computed analytically; 
for a square pulse with length $L$, assuming that the entire pulse is converted at the phase-matched frequency (and so $L/Z_0 = \pi/2$), the crosstalk is
\begin{equation}
\chi = \sin^2 \left(\frac{\pi}{2} \mathrm{sinc}\left(\Delta q L / 2\right) \right) ~.
\label{chi_square}
\end{equation}

For a Gaussian pulse with an interaction length of $L$ the crosstalk is 
\begin{equation}
\chi = \sin^2 \left(\frac{\pi}{2} \exp\left(-1/2 (\Delta q L /2)^2 \right)  \right)~.
\label{chi_gauss}
\end{equation}

\begin{figure}
   \centering
   \includegraphics[scale=0.46]{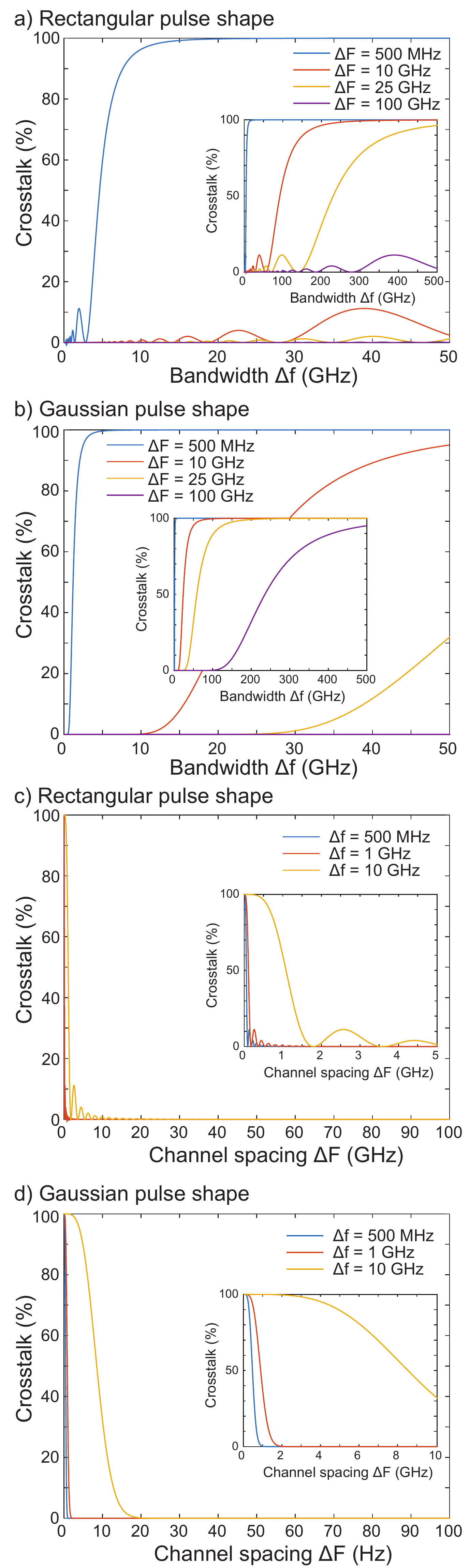}
 \caption{Crosstalk study: a) and b) Fixed channel spacing $\Delta F$ for varying bandwidth $\Delta f$; c) and d) fixed bandwidth $\Delta f$ and varying channel spacing $\Delta F$.  The pulse shape is rectangular in a) and c) and Gaussian in b) and d). The respective insets zoom in or out of the x-axis with reasonable bandwidth and channel spacing for the purpose of demonstrating the behaviour of the crosstalk outside these practical values.}
 \label{fig9}
\end{figure}

The phase-mismatch increases proportionally with the difference in frequency $\Delta F$ from the phase-matching point via
Eq. (\ref{DeltaqEq}). At the same time, for a sufficiently long waveguide the interaction length will be entirely determined by the pulse length, which is inversely proportional to the spectral width $\Delta f$ of the pulse via
\begin{equation}
L = \frac{c \sigma}{n_{\rm eff} \Delta f}~,
\label{Leq}
\end{equation}
where $\sigma$ is the pulse shape parameter, which is equal to 0.8859 for a square pulse 
(arising from the solution to the transcendental equation $\mathrm{sinc} (\sigma \pi / 2) = 1/\sqrt{2}$), and is equal to $2 \ln 2 / \pi \approx 0.4413$ for a Gaussian pulse.
Combining Eqs. (\ref{DeltaqEq}) and (\ref{Leq}) we obtain 
\begin{equation}
\frac{\Delta q L}{2} = 2 \pi \sigma \frac{\Delta F}{\Delta f}~.
\label{dqLdf}
\end{equation}
The crosstalk is therefore determined by the square of the ratio between the pulse linewidth and the distance in frequency
between the pulses. From Eq. (\ref{chi_square}) the crosstalk for a square pulse falls off as a $\rm sinc$ function, with the maximum crosstalk occurring at regular peaks with values $\chi_{\rm max} \sim \pi^2 / (\Delta q L)^2$ 
for large values of $\Delta q L$. Combining with Eq. (\ref{dqLdf}), the peak crosstalk therefore scales as
\begin{equation}
\chi_{\rm max} \sim \frac{1}{(4\sigma)^2} \left( \frac{\Delta f}{\Delta F} \right)^2 ~.
\label{crosstalk1}
\end{equation}
This reduction in crosstalk corresponds to a factor of $\sim 60$ dB for pulses with pulse length of 10\,cm and separated by 100\,GHz. 
For a Gaussian pulse, the crosstalk falls off exponentially:
\begin{equation}
\chi \sim \left(\frac{\pi}{2}\right)^2 \exp\left[-(2 \pi \sigma \Delta F / \Delta f)^2 \right] ~.
\label{crosstalk2}
\end{equation}
Both expressions (\ref{crosstalk1}) and (\ref{crosstalk2}) are independent of the refractive index and acoustic properties of the waveguide, reflecting the fact that the reduced crosstalk arises from the phase mismatch between the modes.
\\
Equations \ref{chi_square} and \ref{chi_gauss} for the respective crosstalk are plotted in Fig.~\ref{fig9}, in one case for a fixed channel spacing between two wavelength channels and in the other case for a fixed bandwidth of the data pulses. The crosstalk is plotted for pulses with a bandwidth up to 50\,GHz for four different fixed channel spacings, 500\,MH, 10\,GH, 25\,GH, and 100\,GHz, in Fig.~\ref{fig9}a and b. For a rectangular pulse shape (Fig.~\ref{fig9}a), significant crosstalk (>\,5$\%$) can be expected at a channel spacing smaller than 10\,GHz and a pulse bandwidth above 20\,GHz. Closely spaced channels of only 500\,MHz spacing lead to high crosstalk already for 1GHz-pulses. For Gaussian pulses, crosstalk increases faster with increasing bandwidth of the pulses, i.e. for a 10\,GHz channel spacing, significant crosstalk is expected at about 15\,GHz pulse bandwidth (Fig.~\ref{fig9}b). The insets in Fig.~\ref{fig9}a and b show the behaviour of the crosstalk in case of an unreasonable high pulse bandwidth up to 500\,GHz, for the purpose of illustration. The crosstalk for a fixed pulse bandwidth of 500\,MHz, 1\,GHz and 10\,GHz, for a varying channel spacing up to 100\,GHz is shown in Fig.~\ref{fig9}c and d. The crosstalk decreases rapidly for reasonable channel spacing values. For Gaussian pulses with a 10\,GHz bandwidth, one has to ensure that the channel spacing is at least 20\,GHz. The insets of Fig.~\ref{fig9}c and d show a zoom into closer channel spacing which can be achievable with adequate experimental equipment. For 1\,GHz pulse bandwidth one can chose a channel spacing as close as 1\, GHz (rectangular shape) or 2\,GHz (Gaussian shape) without considerable crosstalk.

\section{Conclusion}

In conclusion, we have experimentally demonstrated that the accumulated phase mismatch allows to coherently transfer optical information to two acoustic modes even if their frequency spacing is less than their intrinsic linewidth. This finding prevents crosstalk between opto-acoustic processes at close acoustic frequencies, while spatially and temporarily overlapping in the waveguide. The wavelength of the interacting optical waves is preserved which allows for parallel signal processing. In the specific case of Brillouin-based memory, this feature allows for simultaneous storage at multiple frequencies and therefore greatly enhances the capacity of this storage scheme. It offers the possibility to selectively delay data in a specific wavelength channel without effect on other wavelength channels. We have also experimentally shown that the coherence is maintained and not affected by parallel opto-acoustic interaction at another wavelength. 

From a fundamental point of view, this demonstration is an essential step for further investigating the interaction between different acoustic phonons. As no crosstalk from the optical waves is expected, a possible interaction between acoustic modes at different frequencies, located at the same position, could potentially be observed, specifically in a nonlinear acoustic regime.

Selective opto-acoustic interactions at multiple wavelengths in the backward SBS process has far-reaching consequences beyond the previously described light storage, e.g. Brillouin sensing, frequency combs, microwave processing, multi-wavelength Brillouin lasers and quantum interactions. Being suitable for wavelength division multiplexing, future multi-wavelength sensing schemes could benefit in terms of sensitivity enhancement, sensing speed and the discrimination of different sensing parameters such as temperature and strain. It can extend existing microwave processing schemes, e.g. versatile filtering and phase shifting, to multiple wavelengths while avoiding detrimental crosstalk and signal degradation. One can also imagine to process frequency combs selectively~\cite{Tkach1989} and simultaneously coupling individual comb lines to different acoustic modes.

The findings presented here also illustrate a subtle point about the nature of phonon modes in waveguides, which is that these modes remain distinct entities even when they overlap considerably in the frequency domain. This is explicitly demonstrated by the coherent storage of information in phonon fields that are not only indistinguishable in their spectra but also overlap in the spatial domain. The storage process therefore involves distinct phonons at the fundamental level, as opposed to a spectrally-broadened coupling to a single specific phonon field. This has implications for the modelling and understanding of multi-order cascaded Brillouin lasers~\cite{Braje2009,Buettner2014a,Buttner2014a,Merklein2015,Iezzi2016,Gundavarapu2017}, in which it becomes evident that multiple independent amplitudes must be used to represent the phonon fields.

Multiplexing the frequency is also an important feature for quantum interactions and quantum state transfer as this enhances the capacity for parallel quantum communication channels as for example shown in the case of an atomic quantum memory using wavevector multiplexing~\cite{Parniak2017}. The phase matching in SBS ensures also that no scattering into different wavelength channels occurs which ultimately reduces the noise.

As the strong coupling regime is within reach in SBS devices~\cite{VanLaer2016,Kien2016}, efficient transfer at the quantum level might be possible. Using a waveguide approach for storage has the unique advantage of being frequency preserving in a certain frequency window and enables parallel processing at multiple optical wavelength channels, which distinguishes this concept from many opto-mechanical cavity approaches where instead frequency conversion is possible~\cite{Hill2012}.

\section*{Acknowledgements}
This work was sponsored by the Australian Research Council (ARC) Laureate Fellowship (FL120100029) and the Centre of Excellence program (CUDOS CE110001018). We acknowledge the support of the ANFF ACT.

\end{document}